\documentclass[preprint]{aastex6}
\pdfoutput=1
\usepackage{graphicx}
\usepackage{natbib}
\usepackage{amssymb}
\usepackage{amsmath}
\usepackage{float}
\begin{document}
\title{Magnetic Flux Emergence and Decay Rates for Preceder and Follower Sunspots Observed with HMI}
\author{A.A. Norton\altaffilmark{1}}
\affil{HEPL, Stanford University, Stanford, CA 94305, USA}
\author{E.H. Jones\altaffilmark{2}}
\affil{The University of Southern Queensland, Toowoomba, QLD 4350, Australia}
\author{M.G. Linton\altaffilmark{3}}
\affil{US Naval Research Laboratory, 4555 Overlook Ave., SW Washington, DC, 20375, USA}
\author{and J.E. Leake\altaffilmark{4}}
\affil{NASA Goddard Space Flight Center, 8800 Greenbelt Rd., Greenbelt, MD, 20771, USA}

\begin{abstract}
We quantify the emergence and decay rates of preceder (p) and follower (f) sunspots within ten active regions from 2010-2014 using Space-weather Helioseismic Magnetic Imager Active Region Patch data. The sunspots are small-- to mid-sized regions and contain a signed flux within a single polarity sunspot of $(1.1-6.5)\times 10^{21}$ Mx. The net unsigned flux within the regions, including plage, ranges from $(5.1-20)\times 10^{21}$ Mx. Rates are calculated with and without intensity contours to differentiate between sunspot formation and flux emergence. Signed flux emergence rates, calculated with intensity contours, for the p (f) spots average $6.8$ (4.9) $\times 10^{19}$ Mx h$^{-1}$, while decay rates are $-1.9 (-3.4)\times 10^{19}$ Mx h$^{-1}$. The mean, signed flux emergence rate of the regions, including plage, is $7.1 \times 10^{19}$ Mx h$^{-1}$ for a mean peak flux of $5.9 \times 10^{21}$ Mx. Using a synthesis of these results and others reported previously, there is a clear trend for larger flux regions to emerge faster than smaller ones. Observed emergence rates ($d{\phi}/dt$, Mx h$^{-1}$) scale with total signed peak flux, $\tilde{\phi}_{max}$, as a power law  with an exponent of 0.36, i.e., $d{\phi}/dt = A \tilde{\phi}_{max}^{0.36}$. The observed rates may assist in constraining the boundary and initial conditions in simulations which already demonstrate increased rates for flux tubes with higher buoyancy and twist, or in the presence of a strong upflow. Overall, the observed emergence rates are smaller than those in simulations, which may indicate a slower rise of the flux in the  interior than captured in simulations.
  \end{abstract}

\section{Introduction}

Sunspots are magnetic features observed in the photosphere whose properties over a solar cycle seem to imply the presence of a dynamo located somewhere within the interior of the Sun (see for example \citealp{babcock:1959,charbonneau:2005,brandenburg:2005,fan:2009a,stein:2012,yeates:2013}), and no small effort has been exerted in exploring the short term characteristics of sunspot groups as well as their long-term aggregate behavior, in both the observational and numerical realms (see for example \citealp{ferriz-mas:1990,title:1993,miesch:2005,rubio:2008,rempel:2011,toriumi:2013,ilonidis:2011}).

Sunspots are dark due to the partial suppression of convection by a strong (kG range) magnetic field (see for example \citealp{schluter:1958,thomas:2008,priest:2014} and references therein). Their properties over a solar cycle are used to establish many of the characteristics of the Sun that numerical models of the solar interior seek to reproduce \citep{caligari:1995,charbonneau:2005}. The origin of the magnetic flux that gives rise to sunspot groups is not entirely understood. Most models of the solar interior place its source in a dynamo at the base of the convection zone near the tachocline, where turbulent pumping into a subadiabatic overshoot layer can store and amplify magnetic flux \citep{insertis:1992, insertis:1995,charbonneau:2005,ferriz-mas:2007}. However, it should be mentioned that some models have shown that it is possible to form strong magnetic fields within the bulk of the convection zone, a process described by the distributed, or near-surface dynamo \citep{brandenburg:2005,stein:2012} and also through global dynamo models capable of generating buoyant flux loops \citep{nelson:2011, nelson:2014}. There are many problems present in the paradigm of the near-surface dynamo scenario, mainly that Hale's law -- that the ordering of the group magnetic polarity in the East-West direction is opposite in the North and South hemispheres for a given sunspot cycle and that this ordering changes from one cycle to the next -- is not explained nor is the incoherence of the follower sunspot \citep{brandenburg:2005,norton:2013}. Theoretical explanations for the East-West asymmetries of sunspots have been reported from simulations of flux tubes rising from near the base of the convection zone. For example, \citet{dsilva:1993} and \citet{caligari:1995} report that the tilt angles of active regions are reproduced while \citet{fan:1993} show that the incoherence of follower spots can be simulated.

Much research has been devoted to sunspot emergence and decay.  The vast majority analyzed photometric data and reported rates in units of microhemisphere, $\mu$Hem, per time since white-light recordings of sunspots have the longest observational history.  More reports are available on decay than on emergence, probably because decay occurs on a slower timescale so it is more likely to be fully observed. While we cite a few photometric studies, we are particularly interested in the more recent studies analyzing line-of-sight magnetogram or vector magnetic field data with a cadence of hourly or better so that we can compare magnetic flux values measured with the Helioseismic Magnetic Imager (HMI) on board the Solar Dynamics Observatory (SDO) \citep{scherrer:2012}. 

SDO/HMI is a filtergram instrument that images the full disk of the Sun at the Fe-I 6173 \text{\AA} absorption line with a pixel size of  $\approx$0.5$^{\prime\prime}$ pixel$^{-1}$ using a 4096$\times$4096 CCD \citep{schou:2012,scherrer:2012}.  The filtergram images are recorded for six wavelength positions across the spectral line in a combination of polarization states to acquire the Stokes I, Q, U and V data.  The HMI team produces observables of Doppler velocities, line-of-sight magnetic field values, line widths, line depths, and intensities every forty-five seconds while providing vector magnetic field quantities derived from the Very Fast Inversion of the Stokes Vector (VFISV) code every twelve minutes \citep{metcalf:1994,borrero:2011,centeno:2014, hoeksema:2014}. 

For this paper, we identified active regions observed by HMI that fully emerged on the visible portion of the solar disk in order to quantify the flux emergence rates within umbra, umbra plus penumbra, then umbra plus penumbra plus the local region surrounding the sunspots. We separated the magnetic polarities in order to detect differences between the preceder (p) and follower (f) sunspot behavior. The active regions began to decay before rotating off the West limb but we do not sample all of the decay phase. Because we prioritized the emergence over the decay process, the decay rates quoted in this paper are not as robust as the emergence rates. We quantify the percentage of flux contained within the umbra, the percentage contained in the penumbra and umbra, the maximum separation in Megameters (Mm) between the f and p spots, and how the flux emergence and decay rates depend on peak flux. The results of our study are put into perspective by reviewing numerous observations and numerical simulations. 

{\bf \em Why is the flux emergence process interesting?} Flux emergence triggers transient events such as flares, coronal mass ejections and jets. Helicity is transported through the emergence and eruption of magnetic structures.  Comparing the observed flux emergence process to numerical studies allows us to determine if the physical processes included in the simulations are realistic and capturing the dominant solar processes. 

{\bf \em How does this paper add to the existing literature on observed flux emergence?}  This study adds an HMI vector magnetogram based study to the growing number of reports of flux emergence and decay rates.  Our work is similar to \cite{Toriumi:2014} in that they used HMI to study active regions. However, they use the line-of-sight magnetogram data instead of the vector magnetic field data and they report on the maximum flux emergence rates while we report the average rate. \cite{Toriumi:2014} also did not report on the p and f polarities or cite decay rates but focused on velocity dynamics of the emerging region. \cite{otsuji:2011} presents an impressive study with one hundred and one flux emergence events using filtergrams from the Solar Optical Telescope \citep[SOT,][]{tsuneta:2008} 
aboard Hinode \citep{kosugi:2007}. However, most of those events are observed on the order of hours and study only the line-of-sight magnetic field whereas we follow the emerging flux regions for a week on average and use vector field data. Our observed rates of active region flux emergence are shown in context with other reported values, including ephemeral region emergence and simulation emergence rates.  We examine whether regions containing less flux exhibit slower emergence rates.  \textbf {The aims of this paper are to quantify the magnetic flux emergence and decay rates of active regions using vector magnetic field data in order to constrain the simulation conditions and understand the subsurface emergence process that we cannot observe directly.}

HMI is well-suited for capturing bulk properties of emerging flux regions such as the average emergence rate. Since HMI is a full-disk imager with nearly continuous coverage, one can track the region with a varying bounding box during its full rotation across the disk, and construct a union of these boxes to determine the maximum size of the region. The small-scale processes that need higher spatial resolution and more sensitive vector magnetic field measurements, such as submergence of magnetic loops or the presence of bald patches, are better studied with an instrument such as Hinode SOT Spectro-Polarimeter \citep[SP,][]{lites:2013} since it has a higher spatial resolution.  However, pointing coordinates must be specified to Hinode SOT SP prior to observations and, as such, the early stages of emergence are often missed or portions of the emergence are captured only for a few hours instead of the entirety. 

 The flux emergence process has been described as a `two-step' process \citep{archontis:2004, otsuji:2010, toriumi:2011}. No doubt there are more than two steps, but it is useful to call attention to an interrupted rise process that occurs in two distinct stages. The first stage being the flux rising through the convection zone due to buoyancy and then stopping just below the atmosphere where the fluid becomes stable to convection. Due to the strongly stratified nature of the convection zone, a fluid element rising from depth expands and spreads into a horizontal sheet \citep{spruit:1987}. This horizontal spreading of the magnetic flux is sometimes referred to informally as `pancaking'.  Near the photosphere, the flux has expanded greatly and dramatically slowed its rise speed in the last few Mm, or halted its rise altogether. The second phase is the emergence of the flux into the atmosphere. It is still a point of contention whether the magnetic flux must then experience another magnetic buoyancy instability \citep{acheson:1979, archontis:2004} to emerge into the solar atmosphere or whether convective motions push it through into the atmosphere. Simulations show that parameters such as the axial field strength, twist and density of the flux tube dictate whether the flux can fully or partially emerge into the solar atmosphere \citep{magara:2006, murray:2006, cheung:2007, toriumi:2011}. It is not known how much flux is trapped in the near-surface layers.  

While our study is limited to the photosphere, others have observed the rise of flux through different layers of the atmosphere, notably \citet{Ortiz:2014} who studied granular sized magnetic bubbles and \citet{otsuji:2010} who report on the early stages of an emerging active region and associated arched filament system. Both report intermittent expansion of flux systems in the photosphere prior to a rise into the upper photosphere and chromosphere, thus supporting a two-step process in the atmosphere where magnetic plasma piles up and flattens prior to developing an instability that allows it to rise further.


Some details of flux emergence rates in the photosphere, most often reported in units of Maxwells per hour (Mx hr$^{-1}$), observed by others are as follows. 
\begin{itemize}
\item \cite{otsuji:2011} reported on one hundred and one flux emergent events observed with Hinode SOT. The events were followed in time for one to twenty-four hours, meaning that they were studied for usually much less than one day. An average, signed flux  emergence rate of 2.99 $\times 10^{19}$ Mx hr$^{-1}$ was reported for a wide range of regions with an average, signed flux of 4.1 $\times$ $10^{20}$ Mx.  Otsuji et al. (2011) found a power law relationship where average flux emergence rates, $d{\phi}/dt$ in units of Mx hr$^{-1}$, were dependent on peak flux, $\phi$ in units of Mx, in the following form,  $d{\phi}/dt = 9.6\times 10^{7} \phi^{0.57}$. The units of the constant, $9.6\times 10^{7}$, were not stated but presumably must be in Mx hr$^{-1}$ and $\phi$ effectively divided by a normalizing constant of 1 Mx. 

\item \cite{Toriumi:2014} studied twenty-one active regions and two ephemeral regions using HMI line-of-sight magnetic and Doppler data, reporting on strong horizontal diverging flow fields, the peak flux emergence rates and footpoint separations.  The regions had a net magnetic flux that ranged between $0.56 - 23.0 \times $10$^{21}$ Mx with peak emergence rates of $1.3-17.0 \times $10$^{16}$ Mx s$^{-1}$ (equivalent to $4.7-61.2 \times $10$^{19}$ Mx hr$^{-1}$). We report values for a single magnetic polarity within this manuscript, so divide those of \cite{Toriumi:2014} by two for comparison.

\item Using HMI vector magnetic field data for two emerging regions, \cite{centeno:2012} found a rate of $4 \times $10$^{19}$ Mx hr$^{-1}$ for NOAA 11105 with peak flux of $1.4 \times $10$^{22}$ Mx, and a rate of $\approx1.6 \times $10$^{19}$ Mx hr$^{-1}$ for NOAA 11211 with a peak flux of $1\times 10$$^{21}$ Mx. 
The peak, unsigned flux values for those active regions were not reported in \cite{centeno:2012}  but were determined from HMI data.

\item \cite{rezaei:2012} studied the penumbral formation using the German Vacuum Tower in Tenerife (VTT) and using Tenerife Infrared Polarimeter (TIP) and found that within a 4.5 hour observing window, the magnetic flux of the penumbra increased at a rate of $1.89 \times $10$^{20}$ Mx hr$^{-1}$ while the magnetic flux of the umbra remained constant at 3.8x10$^{20}$ Mx.  \citet{schlichenmaier:2010a} and \citet{schlichenmaier:2010b} reported on a penumbra forming at a rate of $1.4 \times $10$^{20}$ Mx hr$^{-1}$ for a sunspot of 1 $\times$ 10$^{22}$ Mx and reported that during the penumbral formation, no pore merged with the sunspot, but rather the increase in magnetic flux was due exclusively to the emergence of small-scaled bipoles.  While it may not be standard to include penumbral formation as a flux emergence rate, these penumbral studies show that the penumbra forms quickly due exclusively to small-scale flux emergence incorporated into the penumbra within hours.

\item Regarding the emergence rates of ephemeral regions, \cite{hagenaar:2001} reports a net, unsigned magnetic emergence rate of $1.1\times$ 10$^{19}$ Mx hr$^{-1}$. Another study of ephemeral regions was carried out by \cite{yang:2014} who found a flux emergence rate of 1 $\times$10$^{19}$ Mx hr$^{-1}$. We report signed flux values within this manuscript and therefore halve those of \cite{hagenaar:2001} and \cite{yang:2014} to compare.

\item The photometric study of \cite{dalla:2008}, using the USAF / Mt Wilson catalogue, reported an average emergence rate of 30 to 40 microHemispheres ($\mu$Hem) per day for sunspots which corresponds to an average rate of 0.8 to 1.1$\times$10$^{20}$ Mx hr$^{-1}$. 

\item \citet{schrijver:2000b} studied full-disk emergence rates and found that they range from $3 \times 10$$^{19}$ Mx hr$^{-1}$ at solar minimum to $2.5 \times $10$^{20}$ Mx hr$^{-1}$ at maximum \citep{schrijver:2000b}.  These full-disk values represent a global flux budget and do not represent individual emerging flux regions.  
\end{itemize}

We now summarize a few representative flux emergence simulation studies for comparison with these observational efforts. This is a very limited list meant to put the observations in context and is not exhaustive by any means. The text of these papers were chosen because they easily provided flux emergence rates for comparison with observations or because these rates could be provided to us by co-authors on this paper.
\begin{itemize}
\item \cite{cheung:2007} simulated flux emergence in convective granules running magnetohydrodynamic simulations with the MURaM code \citep{vogler:2005} with flux tubes containing up to $10^{19}$ Mx. They reported on eight simulation runs for the flux emerging in a twenty minute period with rates that ranged from $1.35 - 16.7 \times 10^{19}$ Mx hr$^{-1}$ depending on field strength at the tube axis, total longitudinal flux and twist. The simulation domain was $24 \times 12 \times 2.3$ Mm$^{3}$.  The top boundary is chosen to be 450 km above the photosphere, defined as the mean geometrical height where the continuum optical depth at 500 nm is unity.

\item \cite{cheung:2008} performed two runs of simulated emerging flux in the presence of convection. Using a flux tube with initial net (unsigned) flux at the base of the simulation of $1.55 \times $10$^{20}$ Mx, they found an emergence rate of $3 \times $10$^{19}$~Mx hr$^{-1}$ for a twisted tube, and half that for not twisted. The simulation domain was 24~$\times$~18 $\times~$5.76 Mm$^3$.  The top boundary is chosen to be 300 km (two pressure scale heights) above the mean geometrical height where the continuum optical depth at 500 nm is unity.

\item \cite{rempel:2014} used a larger domain ($\approx$147$\times$74$\times$16 Mm$^3$) to simulate the emergence and decay of an active region in a fully convective medium. A semitorus-shaped flux tube was advected across the bottom boundary of the simulation. They found that a flux loop containing $1.7 \times $10$^{22}$ Mx has an emergence rate of $6.6 \times $10$^{20}$ Mx hr$^{-1}$, which is about 10$\times$ faster than \cite{centeno:2012} observed for a similar-sized (total flux) active region emergence.  The top boundary is chosen to be approximately 700 km above where continuum optical depth at 500 nm is unity. 

\item \citet{Leake:2016} used a domain of 87 Mm in all directions, including the upper 34 Mm of the convection zone, simulating the emergence of magnetic flux tubes from the convection zone through the photosphere and chromosphere and into the corona to a height of 53 Mm above the photosphere. Their simulations did not include pre-existing convection but did include the interplay of buoyancy and buoyancy-driven convective motions  in the convection zone, as well as the later emergence of flux into the corona. They varied the parameters of the initial flux tube and found flux emergence rates of [0.9, 4 and 9] $\times $10$^{19}$~Mx hr$^{-1}$ for active regions of flux [0.4, 3 and 10] $\times $10$^{19}$~Mx, respectively.
\end{itemize}

There are a few aspects to note when comparing observations of flux emergence to numerical models. First, not all of the flux in a simulation emerges into the atmosphere; some stays trapped below the surface.  Second, the amount of flux crossing the photospheric surface can be higher than the flux contained within the flux tube since the field becomes serpentine, threading in and out of the surface. This process can cause the amount of flux to be up to an order of magnitude higher than the original longitudinal flux in the tube, see Table 1 in \cite{cheung:2007}. 

For a comprehensive review of flux emergence theory, see \cite{cheung:lrsp}. For a review that features both observations and theory, see \cite{archontis:2012}.  This study, focusing on a select group of ten sunspots in Solar Cycle 24, observed with HMI, will provide observational guidance for physical interpretation and numerical simulations of magnetic flux emergence from the converction zone into the solar atmosphere.  
\begin{table}[h!]
\footnotesize
\centering
\begin{tabular}{cccccccccc}
\hline\hline
 NOAA & HARP & Emergence (UT) & Observation End (UT) & Hem & Lat & Joy\\ 
\hline
11141 & 325 &2010 Dec 30 21:47 &2011 Jan 04 01:47 & N &  34$^o$ & Y\\ 

11184 & 466 &2011 Mar 31 13:47 &2011 Apr 07 12:59 & N & 15$^o$  & Y\\ 

11428 & 1447 &2012 Mar 02 23:11 &2012 Mar 11 18:35 & S &  -17$^o$ & Y\\ 

11460 & 1578 &2012 Apr 15 11:47 &2012 Apr 23 18:23 & N &  15$^o$ & Y\\ 

11465 & 1596 &2012 Apr 19 14:59 &2012 Apr 28 07:23 & S &  -17$^o$ & N\\ 

11497 & 1727 &2012 May 31 12:47 &2012 Jun 08 17:11 & S &  -21$^o$ & Y\\ 

11512 & 1795 &2012 Jun 24 21:35 &2012 Jul 01 16:59 & S &  -16$^o$ & Y\\ 

11682 & 2501 &2013 Feb 25 00:47 &2013 Mar 02 22:11 & S &  -19$^o$ & N\\ 

11723 & 2663 &2013 Apr 13 04:23 &2013 Apr 21 11:27 & S &  -19$^o$ & Y\\ 

12053 & 4092 &2014 May 03 05:35 &2014 May 12 06:35 & N &  10$^o$ & Y\\ 

\hline
\end{tabular}
\caption{\small Active regions examined in this paper. Listed are: NOAA Active Region number, HMI Active Region Patch number, time of initial pore formation in Universal Time (UT), observation end time, solar hemisphere, approximate latitude and whether or not the active region obeys Joy's law. Note that cutoff times are based on either the complete decay of the sunspot region or its proximity to the solar limb. NOAA 11465 and NOAA 11682 both displayed tilt angles during their evolution that were anti-Joy, in that their follower sunspots were on average closer to the equator than that of the preceder ones.}
\label{table-ars}
\end{table}

While we emphasize the flux emergence process in this paper, it was easy to compute the decay rates of the ARs as well. Interpreting the decay rates is another matter, altogether, and not as easily done. A much-studied aspect of decaying sunspots is the moving magnetic feature (MMF). MMFs are small knots, roughly $10^3$ km in diameter, of magnetic flux in the range of $10^{18}$ $-$ $10^{19}$ Mx that move outward from the parent spot at about 1 km s$^{-1}$ \citep{sheeley:69, hh:73, hagenaar:2005}. At first glance, it would appear that these features are responsible for sunspot decay. However, they are observed to transport flux away from spots at a higher rate than the observed flux loss \citep{hh:73, pillet:2002}.  Statistical properties of MMFs observed with the Michelson Doppler Imager were reported by \citet{hagenaar:2005} who found flux transport rates away from the parent sunspots of $(0.4-6.2) \times 10^{19}$ Mx h$^{-1}$. \citet{kubo:2008} reports an actual flux loss of 3.9 $\times$ 10$^{15}$ Mx s$^{-1}$ ($1.4 \times 10^{19}$ Mx h$^{-1}$) from a sunspot observed with Hinode SOT while the flux carried across a circular boundary 20$\arcsec$ away from the sunspot center by MMFs is almost double that.  MMFs are observed to cancel with opposite polarity flux as a way of removing the flux from the photosphere \citep{kubo:2008}. We know of no comprehensive study of MMFs using HMI data but \citet{nelson:2016} follow two MMFs with HMI data and find flux cancellation rates of $10^{14}-10^{15}$ Mx s$^{-1}$, or $10^{17}-10^{18}$ Mx hr$^{-1}$. Clearly, MMFs assist in sunspot decay and the removal of flux from the photosphere but the relative importance of MMFs in relation to other mechanisms is not known. We do not isolate MMFs in this study to determine their role in flux removal from the sunspots, but leave that to future research. We simply report on the decay rates and note that MMFs are clearly present and removing flux, both during the emergence and decay stages. 


\section{Data and Analysis}

We use SDO/HMI continuum intensity and vector magnetogram data from within Spaceweather HMI Active Region Patches (SHARP), which are patch data sets (not full disk) tracked at the Carrington rotation rate and are processed within the HMI pipeline \citep{bobra:2014}.  Data calibration procedures for HMI data are described in \citet{couvidat:2016}.  SHARP data is available in either unprojected (in plate coordinates) form or as Lambert cylindrical equal area (CEA) projected data, see \cite{thompson:2006} and \cite{snyder:1987} for detailed reviews of common solar coordinate systems and cartography. Note that there is no requirement that the net flux within a SHARP region be zero, so there can be an imbalance of flux between polarities \citep{hoeksema:2014}. 
The following criteria were used to select active regions for analysis.  
\begin{enumerate}
\item We define the beginning of active region emergence by the initial appearance of pores in continuum intensity.  Therefore, the initial stages of emergence of the active region as seen in the continuum intensity must occur either entirely or almost entirely on the visible part of the solar disk. We reject regions that rotate onto the disk already having a strong, pore-like decreased intensity.   
\item We require that the group emerge and begin to decay on the visible, Earth-facing side of the Sun. We acknowledge that this criteria selects small- to mid-sized active regions.  Therefore, our results may not apply to the larger active regions.  Also, our study prioritizes emergence over decay, meaning we ensure observation of emergence and do not always capture decay in its entirety.  Therefore, more confidence should be placed on emergence rates reported here and less confidence in the decay rates. 
\item Our study depends upon the existence of preceder and follower sunspots within a group. Therefore, we select only sunspot groups that are bipolar in nature. 
\item The SHARP magnetic field data for a given active region should primarily consist of flux only associated with the region. By this we mean, the flux is not emerging into a region that is already containing a significant amount of flux from a recently emerged or decaying active region.
\end{enumerate}

\begin{figure}[t!]
\centering
\includegraphics[scale=0.95]{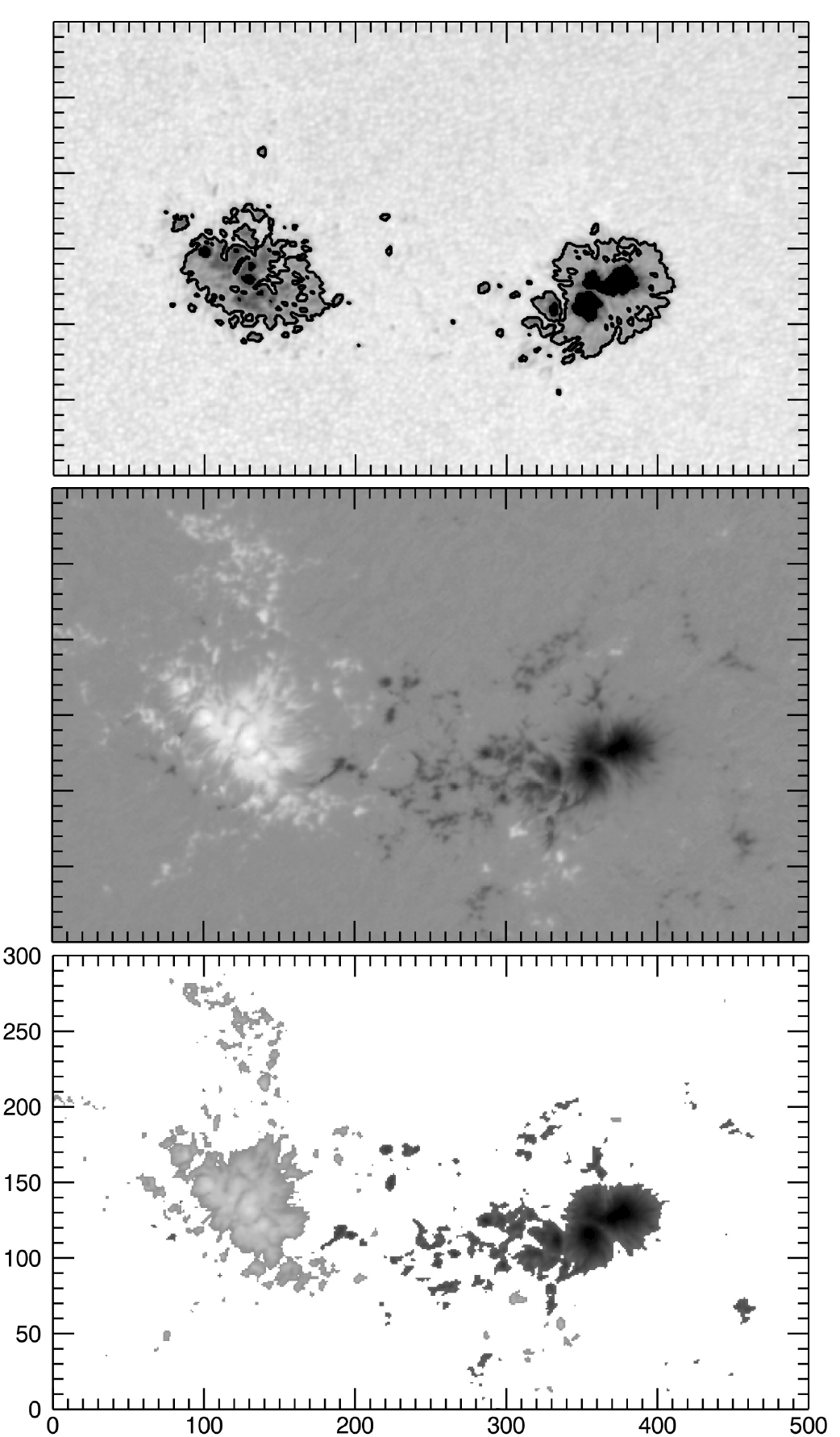}
\caption{\small AR 11460, SHARP data from 2012 Apr 22, is shown in continuum intensity (top), magnetic field normal to the surface, $B_r$ (middle) and the effect of applying a threshold of 575 Mx cm$^{-2}$ such that pixels with $B_r$ values below that strength (shown as white, bottom panel) are not included in the flux total. In the continuum intensity image (top), the penumbral contours are outlined while the umbrae are artificially shaded black. The effect of applying the threshold, as shown in the bottom panel, is that pixels with weak flux, or zero flux, are not included in the temporal analysis of the flux emergence and decay. 
The average flux of the included (not included) pixels was 1081 (67) Mx cm$^{-2}$.  
}
\label{figure-patch-flux-1}
\end{figure}

\begin{figure}[t!]
\centering
\includegraphics[scale=0.70]{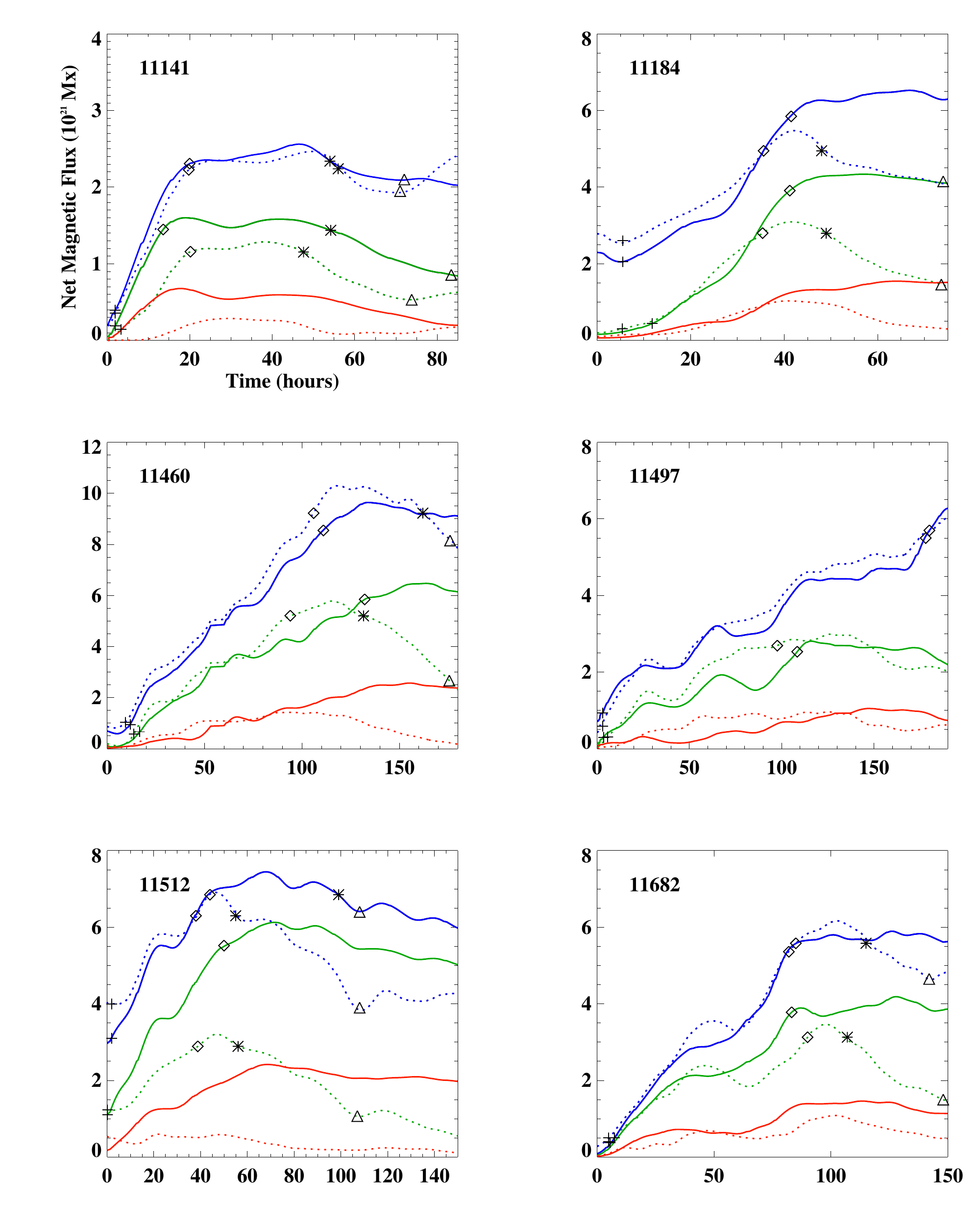}
\caption{\small Net magnetic flux within the preceder and follower regions over time, in units of 10$^{21}$ Mx, from emergence to cutoff time. Solid (dotted) lines represent preceder (follower) flux. The lines are blue for total flux within the SHARP patch, green for flux within the sunspot (umbra plus penumbra), and red for flux within umbrae. We use $+ (\diamond)$ to denote the beginning (end) times used for calculating the emergence rates, and $* (\bigtriangleup)$ to denote the times used for calculating the beginning (end) of the decay rates. Sunspot groups within this series of plots have less than 200 hours of available data.  }
\label{figure-patch-flux-2}
\end{figure}

Table \ref{table-ars} contains general information about the active regions examined in this study, including NOAA number
(\verb;http://www.swpc.noaa.gov/products/solar-region-summary;), HMI Active Region Patch (HARP) number, emergence time which is the beginning of the observation, observation end times, hemisphere, approximate latitude and whether or not the region obeyed Joy's law. Joy's law describes the tendency for the f spot in a sunspot group to be located more poleward, i.e., at a higher latitude, than the p spot and for that poleward displacement to be greater at higher latitudes than near the Equator \citep{hale:1919} although Joy's law has been shown to vary between hemispheres and solar cycle \citep{mcclintock:2013}. All of the active regions exhibited bipolar characteristics and obeyed Hale's law.

\begin{figure}[t!]
\centering
\includegraphics[scale=0.70]{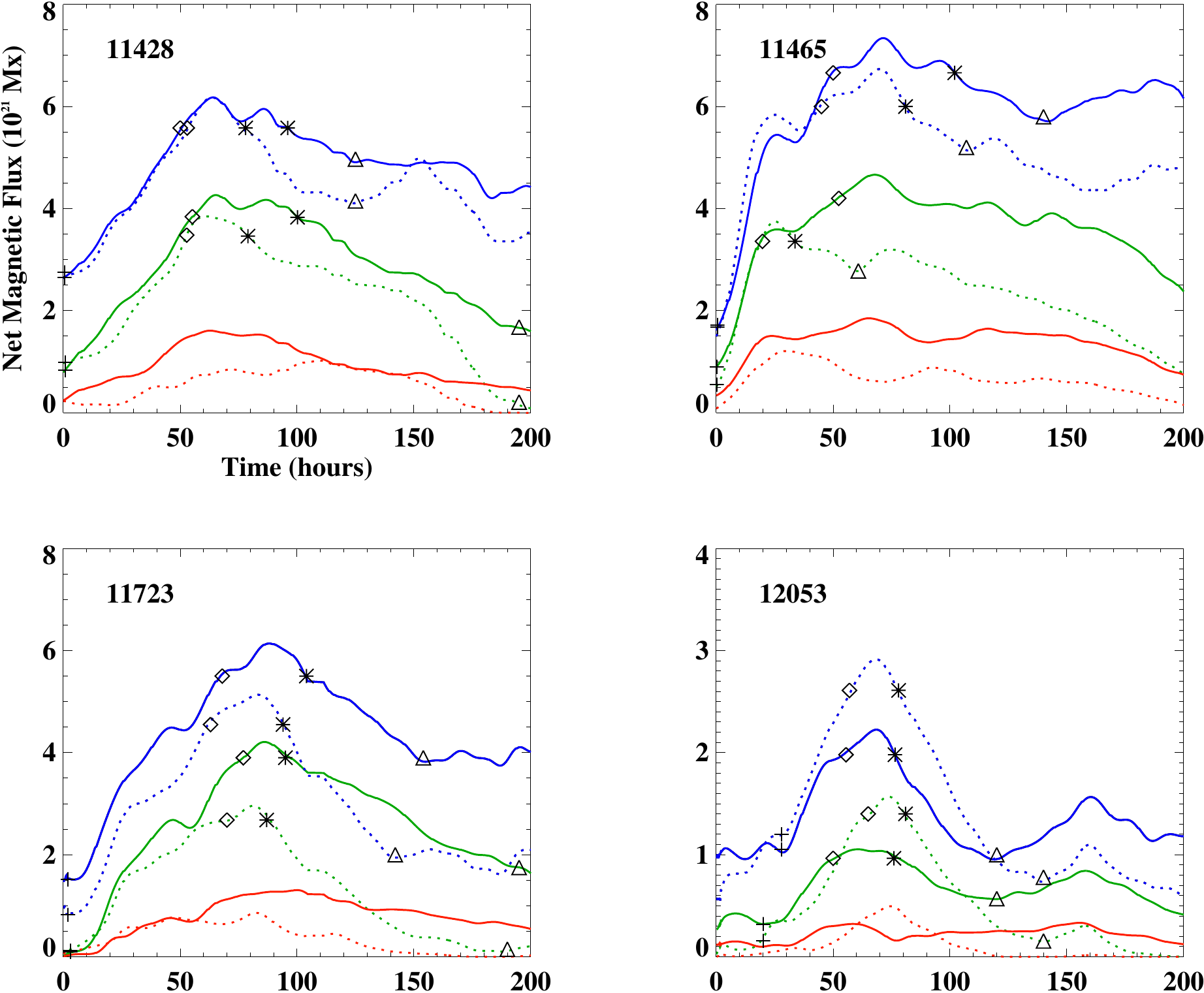}
\caption{\small Net magnetic flux within the preceder and follower regions over time, in units of 10$^{21}$ Mx, from emergence to cutoff time. The lines and symbols have the same meanings as Figure 2. Sunspot groups within this series of plots have more than 200 hours of available data. }
\label{figure-patch-flux-3}
\end{figure}

In order to isolate sunspots from the surrounding photosphere in our data, we first remove limb darkening effects from the continuum intensity data and then determine what thresholds to use on the data in order to define whether a given pixel belongs to the background, a penumbra, or an umbra.  We account for limb darkening by normalizing the continuum intensity within an unprojected patch to a 2nd order polynomial in $\mu$ (cosine of the center-to-limb angle) that has been fit to the quiet sun \citep{hestroffer:1998}. We use a method to threshold the data that is similar to the one used by \citet{verbeeck:2013}, that is, the mean continuum intensity $\bar{I}$ and the corresponding standard deviation $\sigma$ are calculated. We then calculate a threshold intensity, $I_T$ = $\bar{I}$ - $2.58\sigma$. Pixels with a continuum intensity equal to or less than that of $I_T$ belong to a sunspot. To differentiate between the umbrae and the penumbrae, we calculate $\bar{I}_{AR}$ and $\sigma_{AR}$, the mean and standard deviation of just the pixels that belong to sunspots. The threshold for umbra, $I_U$ = $\bar{I}_{AR}$ - $\sigma_{AR}$, is then calculated. Pixels with intensities that lie between $I_T$ and $I_U$ belong to penumbrae, while pixels with intensities below $I_U$ belong to umbrae. As an example, AR 11428 on 2012 March 5 at 07:00 UT has an intensity contrast threshold of 85\% for penumbral pixels and 56\% for umbral pixels in comparison to the quiet Sun continuum intensity.

Our analysis requires that we group sunspots by polarity and calculate magnetic flux. Therefore, we need the component of the magnetic field that is normal to the solar surface. To accomplish this, we use the following SHARP data products: \verb;field;, which gives us $B$, the magnitude of the magnetic field; \verb;azimuth;, which gives us $\phi_b$, the angle between the North direction in plate coordinates and the magnetic field; \verb;inclination;, which gives us $\beta$, the angle between the line-of-sight (LOS) direction and the magnetic field \citep{bobra:2014}. Knowing these quantities, the normal component of the magnetic field is given by
$$ \vec B\cdot \hat r = {{B \over r_s}[d_s-r_s\cos{(\beta + \delta)}-l\cos{\beta}]}$$
\noindent
where $d_s$ is the distance from the center of the Sun to the observer (given by the FITS keyword\verb; DSUN_OBS;), $r_s$ is the radius of the Sun, and $l$ is the distance from the point of interest on the Sun to the observer, and $\delta$ is defined as 
$$\delta = \tan^{-1}\big({r_s\sin\theta\sin(\phi_b-\phi)\over d_s-r_s\cos\theta}\big)$$ 
\noindent
where $\theta$ and $\phi$ are the spherical coordinates of the point of interest before \verb; CROTA2; is applied, with z lying in the direction of the observer and y being North in plate coordinates.

To determine the net magnetic flux within a region, we sum over the associated normal field of each pixel and multiply by the area $A$ of a pixel in CEA coordinates, 
$$\Phi_m = \sum_{i=0}^{n_x}\sum_{j=0}^{n_y} \vec B_{ij} \cdot (A\hat r_{ij}) $$
\noindent
where $n_x$ and $n_y$ are the width and height in pixel units of a patch in CEA coordinates.

\cite{bobra:2014} states that vector magnetogram field strengths that lie below 220 Mx cm$^{-2}$ are generally considered to be noise; we find that there is also a change in the noise in the I, Q, U and V profiles with increasing center-to-limb observing angles, which causes a trend (see Figure 5 in \cite{hoeksema:2014} or Figure 3 in \cite{bobra:2014}) that occurs in the net magnetic flux as the region moves across the solar disk. This trend is due to the VFISV fitting of the lower flux pixels returning stronger transverse field values at large distances from disk center \citep{centeno:2014}. This can translate as a temporal trend as the region rotates nearer the center of the solar disk and can thus confuse the issue of whether a region is emerging or decaying. Short of recalibrating HMI inversion codes, we use a high noise threshold of 575 Mx cm$^{-2}$ which suppresses any disk-crossing trend in the weaker fields. Due to applying this threshold, our reported net flux values are lower than those reported by others who do not threshold, and our flux emergence and decay rates are fractionally lower as well. See Figure 1 for an example of which pixels are included in the analysis of AR 11460 SHARP data based on this threshold.

 Many studies use the line-of-sight magnetic field component ($B_{los}$) instead of the radial magnetic field ($B_r$), as determined from the vector data, to study photospheric flux behavior. $B_r$ has historically been approximated by the $\mu$-correction, i.e. $B_{los}$ divided by the cosine of the observing angle, $\mu$, to approximate $B_r$. The success of this correction is directly related to how radial the structure being examined in the solar atmosphere truly is.  If significant horizontal magnetic field components are present, as are common in active region penumbra, then the $\mu$-correction is not sufficient \citep{leka:2017}.  As noted by \citet{centeno:2014} and \cite{hoeksema:2014}, weak horizontal magnetic field strengths as measured with HMI and inverted with VFISV are problematic near the limb. As such, we choose to disregard measurements in locations more than 70$^{\circ}$ from disk center.

We report the maximum flux of the f and p sunspots, $\Phi_{max}$ in Tables 2 and 3, as a signed quantity. We chose signed, as opposed to unsigned, because it allows us to separate preceder and follower spot behavior, but also because it better relates to the longitudinal flux carried by the flux tube as assigned by the numerical modelers. Peak footpoint separation of the two magnetic polarities was also determined using a center-of-mass method on masked $B_{r}$ data (i.e. flux-weighted method) such that $l_{max}$ is the distance between the positive and negative polarity centroids. 

\begin{table}[t!]
\centering
\begin{tabular}{ccccc|cccc|ccc}
\cline{2-11}
&\multicolumn{4}{c|}{Preceder}&\multicolumn{4}{|c|}{Follower}&\multicolumn{2}{c}{Both}\\
\hline\hline
NOAA & $f_p$ & $\Phi_{max}$ & $\tau_e$ & $\tau_d$ & $f_f$ & $\Phi_{max}$ & $\tau_e$ & $\tau_d$& $l_{max}$ & $r$   \\
\hline
11141	&	0.37	&	1.60	&	10.7	&	-2.1	&	0.18	&	1.29	&	6.0	&	-2.4 & 94.5 &0.81	\\
11184	&	0.34	&	4.34	&	11.7	&	-	&	0.30	&	3.09	&	8.3	&	-4.8	& 80.1 & 0.71\\
11428	&	0.34	&	4.27	&	4.6	&	-2.3	&	0.24	&	3.85	&	5.5	&	-2.8	& 71.6 & 0.90\\
11460	&	0.36	&	6.49	&	4.5	&	-	&	0.24	&	5.78	&	5.8	&	-5.1	& 80.2 & 0.89\\
11465	&	0.38	&	3.75	&	14.4	&	-2.1	&	0.28	&	4.67	&	6.3	&	-        & 53.6 & 1.25\\
11497	&	0.30	&	2.80	&	2.1	&	-	&	0.30	&	2.99	&	2.6	&	-	& 65.8 & 1.07\\
11512	&	0.37	&	6.13	&	8.7	&	-	&	0.18	&	3.21	&	4.3	&	-3.6	& 63.9 & 0.52\\
11682	&	0.34	&	4.19	&	4.4	&	-	&	0.29	&	3.47	&	3.3	&	-4.0	& 71.5 & 0.83\\
11723	&	0.31	&	4.21	&	4.5	&	-2.2   &	0.23	&	3.00	&	3.7	&	-2.5	& 74.2 & 0.71\\
12053	&	0.30	&	1.07	&	2.2	&	-0.9	&	0.21	&	1.55	&	2.8	&	-2.1	& 61.1 & 1.45\\
\hline
\multicolumn{1}{r}{Average}&	0.34	&	3.89	&	6.8	&	-1.90	&	0.25	&	3.29		&4.9	&-3.4	& 71.4 & 0.91\\
\multicolumn{1}{r}{Std Dev}&	0.03	&	1.72	&	4.3	&	0.58	&	0.05	&	1.32	      &1.8& 1.1&11.6	&0.27\\
\end{tabular}
\caption{\small Tabulated results for p and f sunspots (green curves in Figures 1 and 2) within our sample, by NOAA number and with averages and standard deviations shown in the final two rows. The values in this table are calculated using both intensity thresholds and vector magnetic field data with a minimum threshold.  $f_{p}$ ($f_{f}$) is the ratio of flux within the p (f) umbrae (red curves in Figures 1 and 2) to the total flux within the p (f) sunspots, 
$\Phi_{max}$ is the peak flux in 10$^{21}$ Mx, $\tau_e$ is the flux emergence rate in 10$^{19}$ Mx h$^{-1}$, and $\tau_d$ is the flux decay rate in 10$^{19}$ Mx h$^{-1}$.  The use of  '-' indicates that no clear signs of decay were observed. The last two columns contain the peak footpoint separation in Mm between p and f spots, $l_{max}$, and the ratio of f to p flux within spots, $r$. Note that pixels with radial magnetic flux values below a threshold of 575 Mx cm$^{-2}$ are not included in the analysis.} 
\end{table}

There are 24-hour and 12-hour periodicities in strong magnetic fields seen in HMI data that are probably due to a combination of satellite motion and uncertainties in instrument calibration \citep{liu:2012}. We do not attempt to remove these variations as no robust method has yet been developed to remove these variations.  Note, however, that work has recently been done to remove related variations from the Doppler data \citep{schuck:2016}. We do apply a six-hour smoothing function to the flux values for two reasons: to ascertain a unique time for peak flux values and to have a smooth function to determine flux decay periods. The decay period is considered finished when the slope of the flux versus time becomes positive for two consecutive points.  Therefore any increase in flux due to noisy time series would hinder the decay rate determination. SDO/HMI coverage is not always complete. Therefore, we check all data for gaps in coverage, and in instances where gaps exist we de-gap data by first binning the data, and then fitting the binned data to a cubic spline using \verb;FITPACK; routines. To lessen the impact of limb effects, we restrict our analysis to lie between $\pm$ 70$^{\circ}$ of solar center.  The duration of observation for each active region was determined by the location of emergence and how much time HMI could observe the sunspots before they rotated off the West limb. 

In order to calculate emergence and decay rates, we first calculate the peak flux and define points of emergence and decay in reference to the peak flux value. The peak flux, emergence and decay rates are calculated for the following: 
\begin{enumerate}
\item the separated p and f flux with both penumbral and umbral flux included, see Table 2
\item the total flux within the SHARP patch (defined by data bounding box) separated into p and f polarities but without any intensity threshholds being used for calculating if a given pixel was within an umbra or penumbra, see Table 3. 
\end{enumerate}
In all of the above, the flux emergence time period is considered the time between when the flux values reaches ten and ninety percent of the peak values observed.  The flux emergence rate is then simply the amount of flux emerged divided by the time, reported in units of Mx hr$^{-1}$. 

The start of decay was defined as the time after peak flux value at which the feature contained ninety percent of the peak flux value.  The decay was considered finished when the feature either declined to be ten percent of its peak flux value or the slope was positive for two consecutive points in time. 
Some preceder spots did not show decay in our study, such as AR 11184, 11460 and 11682 whose flux values never reduced to ninety percent of the peak value. We also did not determine flux decay rates if the feature did not decay, i.e., have a negative slope, for at least ten hours after it reached ninety percent of its peak flux value. If decay was not observed for ten continuous hours, no decay rate was reported, i.e., AR 11512 p sunspot and AR 11497 p and f spots, see Figure 2.  AR 11465 was also problematic in determining decay rates since new flux emerged in the f spot and the p spot experienced two periods of increasing flux shortly after reaching ninety percent of its peak flux. Therefore, no decay was reported for AR 11465. 

\section{Results \& Discussion}

\begin{table}[t!]
\centering
\begin{tabular}{ccccc|cccc}
\cline{2-9}
&\multicolumn{4}{|c|}{Preceder}&\multicolumn{4}{c|}{Follower}\\
\hline\hline
NOAA & $f_{pp}$ & $\Phi_{max}$ & $\tau_e$ & $\tau_d$ & $f_{fp}$ & $\Phi_{max}$ & $\tau_e$ & $\tau_d$\\
\hline
11141	&	0.61	&	2.64	&	13.9	&	-3.2	&	0.38	&	2.52	&	13.5	&	-2.9	\\
11184	&	0.54	&	6.25	&	10.9	&	-	&	0.44	&	5.42	&	11.6	&	-3.8\\
11428	&	0.57	&	6.21	&	5.3	&	-1.9	&	0.35	&	6.21	&	5.3	&	-3.2\\
11460	&	0.58	&	9.52	&	7.5	&	-	&	0.51	&	10.47&	9.7	&	-3.9\\
11465	&	0.61	&	7.38	&	8.1	&	-3.1	&	0.44	&	6.71	&	7.2	&	-2.7\\
11497	&	0.51	&	5.71	&	2.8	&	-	&	0.53	&	5.49	&	2.8	&	-	\\
11512	&	0.78	&	7.41	&	6.8	&	-1.6	&	0.30	&	6.84	&	4.4	&	-5.1	\\
11682	&	0.71	&	6.10 &	6.7	&	-	&	0.54	&	5.65	&	6.0	&	-4.0	\\
11723	&	0.59	&	6.24	&	5.9	&	-3.8	&	0.26	&	5.21	&	6.1	&	-4.7	\\
12053	&	0.51	&	2.21	&	2.9	&	-2.6	&	0.41	&	2.91	&	4.0	&	-2.8	\\
\hline
\multicolumn{1}{r}{Average}&	0.60	&	5.97	&	7.1	&	-2.7	&	0.42	&	5.74	&	7.1	&	-3.7	\\
\multicolumn{1}{r}{Std Dev}&	0.09	&	2.16	&	3.4	&	0.8	&	0.10	&	2.20	&	3.5	&	0.9	\\
\end{tabular}
\caption{\small Tabulated results for p and f flux within the SHARP patch regions within our sample, by NOAA number and with associated averages and standard deviations. $f_{pp}$ is the ratio of flux within the preceder sunspots to the total preceder flux within the patch. $f_{fp}$ is the ratio of flux within the follower sunspots to the total follower flux within the patch. $\Phi$ is the peak flux within the patch in 10$^{21}$ Mx. $\tau_e$ is the flux emergence rate in 10$^{19}$ Mx h$^{-1}$. $\tau_d$ is the flux decay rate in 10$^{19}$ Mx h$^{-1}$. The use of  '-' indicates that no clear signs of decay were observed. The values in this table are calculated using only vector magnetogram thresholds and not intensity thresholds, except for the $f_{pp}$ and $f_{fp}$ ratios which do use intensity thresholds to calculate the amount of flux within the sunspots. A 575 Mx cm$^{-2}$ threshold is used in that radial magnetic flux values below this are not included in the analysis.} 
\label{table-data-patch}
\end{table}
  Figure 2 shows the flux emergence and decay behavior for six active regions observed for less than two hundred hours. Figure 3 shows the flux emergence and decay behavior for four active regions observed for two hundred hours. The polarities are always separated to show p (solid line) and f (dashed line) flux. In Figures 2 and 3, the behavior is separated into that of flux contained within the umbra (red lines), penumbra and umbra (green lines), and the magnetic flux within the entire patch (blue lines). The magnetic threshold of 575 Mx cm$^{-2}$ is applied to all data plotted within Figures 2 and 3. The intensity thresholds are used to determine flux within the umbra and penumbra but not that within the entire patch, meaning that intensity thresholds are used to calculate the values shown in the red and green curves but not the blue curves.

The resulting emergence and decay rates, as well as peak flux values and ratio of flux within sunspot contours in our sample can be found in Table 2.  p sunspots showed emergence rates that lie between $(2.1-14.4)\times$10$^{19}$ Mx h$^{-1}$, with a mean emergence rate of 6.8$\times$10$^{19}$ Mx h$^{-1}$. However, the standard deviation is quite high at 4.3$\times$10$^{19}$ Mx h$^{-1}$.  Only 50\% of the p sunspots showed clear signs of decay, or satisfied our criteria for decay, with decay rates between $-(0.9-2.3)\times$10$^{19}$ Mx h$^{-1}$, averaging $-1.9\times$10$^{19}$ Mx h$^{-1}$. 

The f sunspots showed a range of emergence rates between $(2.6-8.3)\times$10$^{19}$ Mx h$^{-1}$, with a mean emergence rate of 4.9$\times$10$^{19}$ Mx h$^{-1}$. The standard deviation for f emergence rates is 1.8$\times$10$^{19}$ Mx h$^{-1}$ which is much lower than the standard deviation in the p spots.   80\% of the f sunspots showed clear signs of decay, with decay rates that lie within $-(2.1-5.1)\times$10$^{19}$ Mx h$^{-1}$, averaging $-3.4\times$10$^{19}$ Mx h$^{-1}$. 
This rates compares favorably with flux loss rates by MMFs as reported by \citet{hagenaar:2005} and \citet{kubo:2008} with values of $(0.4-6.2) \times 10^{19}$ Mx h$^{-1}$and $2.8 \times 10^{19}$ Mx h$^{-1}$.

In terms of how compact the p and f spots are, 34\% of the magnetic flux of p spots resides within their umbrae while 25\% of the flux within the f spots resides within their umbrae, as shown in the $f_p$ and $f_f$ values in Table 2.  This supports the well-known fact that follower spots are less spatially coherent, and agrees with observations of fractional flux concentrations made by \cite{zwaan:1992} and the observation of asymmetry in area seen in work done by \cite{murakozy:2012} and others. 

Table 2 also contains some key information about the relationship between the p and f spots such as the peak footpoint separation ($l_{max}$) and the ratio of flux between p to f sunspots ($r$). The average ratio, $r$, of the peak flux in f spots compared to p spots is 0.91. The peak footpoint separation averaged 71.4 Mm $\pm$ 11.6 Mm. We note, with interest, the correlation to the supergranular spatial scale. Power spectra show that supergranular sizes peak around $30-35$ Mm \citep{rieutord:2010}. The maximum footpoint separation shown in Table 2 is consistent with  \citet{mcclintock:2016} who report that active regions of all sizes initially appear as dark features in continuum intensity with a footpoint separation of approximately 40 Mm, then continue to separate over three to four days to a peak footpoint separation of about 70 Mm. The 71.4 Mm value in Table 2 is larger than that reported by  \citep{Toriumi:2014} who found an average maximum footpoint separation of 60.8 Mm for 17 active regions observed with HMI using a flux-weighted centroid method. A systematic study to examine the relationship between supergranule boundaries and initial location of sunspot signatures may prove informative.

Table 3 shows the tabulated results for signed flux in the SHARP patch, not simply within the sunspot. The values in Table 3 are calculated using only vector magnetogram data and not intensity thresholds, excepting to calculate the $f_{fp}$ and $f_{pp}$ values.  The average flux emergence rate is identical for the p and f polarities at $7.1 \times$10$^{19}$ Mx h$^{-1}$.  The decay rates, $\tau_{d}$, show p (f) spots decaying at $-2.7 (3.7) \times$10$^{19}$ Mx h$^{-1}$. A measure of how much of a given polarity flux is contained in plage, as opposed to the sunspot, is found in Table 3 in the $f_{pp}$ ($f_{fp}$) columns. Much of the emergent flux in a given region lies outside of the sunspots umbrae and penumbrae. In the case of preceder polarity, 60\% of the flux actually resides in the sunspots (average $f_{pp}$ value in Table 3) while 42\% of the flux associated with the follower polarity lies in the sunspot (average $f_{fp}$ value in Table 3). 

\begin{figure}[t!]
\centering
\includegraphics[scale=0.80]{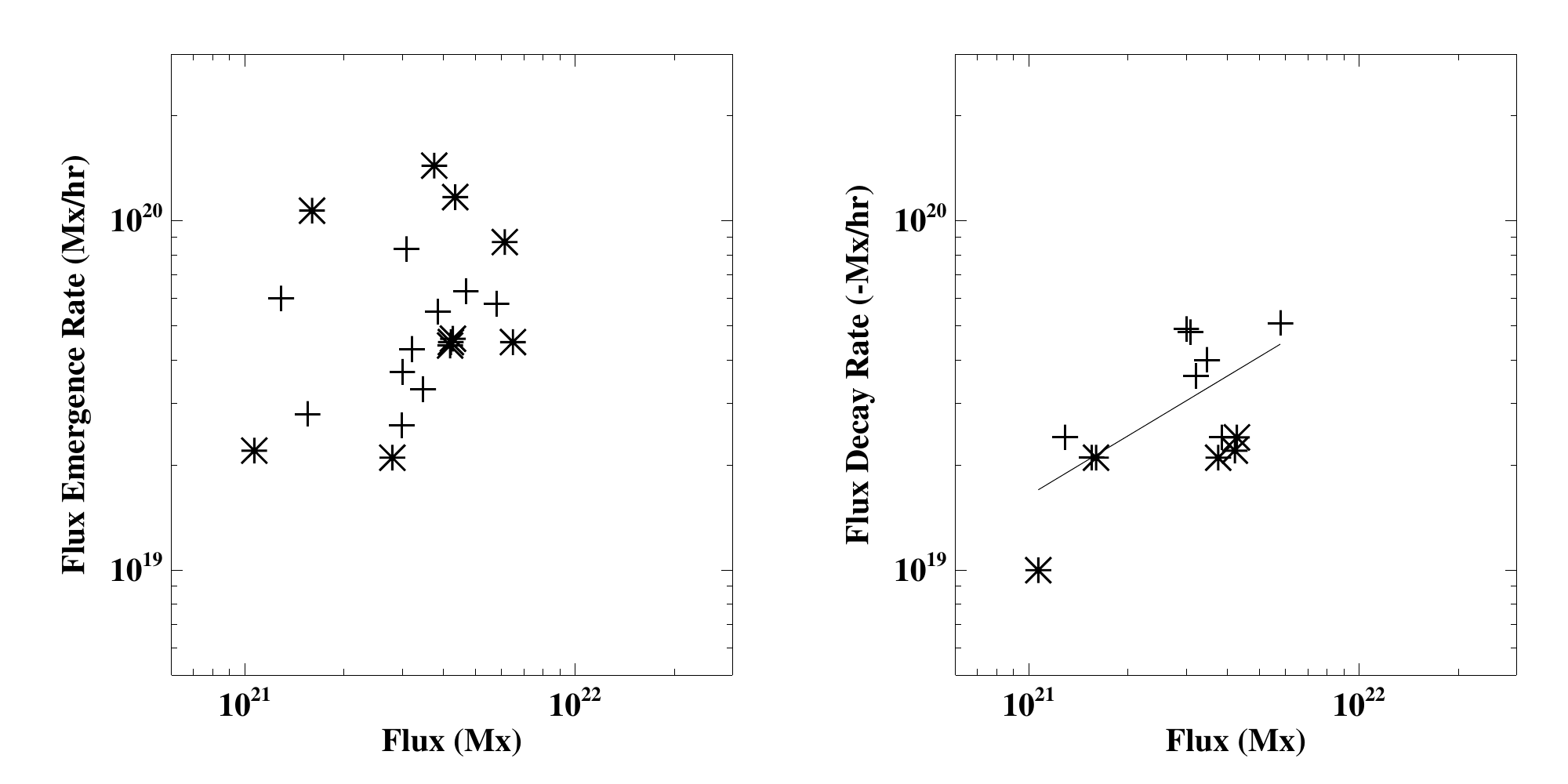}
\caption{Data from Table 2 is plotted in log-log format as emergence (left) and decay (right) rates versus the net unsigned peak flux for the p (*) and f (+) polarities within our active region sample.  Emergence rates were not correlated with peak flux. Decay rates were better correlated with peak flux with linear Pearson correlation coefficients of 0.78 (0.63) for p (f) spots, and scaled as a power law with exponent of 0.57 so that $- d{\phi}/dt \propto {\phi}_{max}^{0.57}$. } 
\label{figure-area-decayplots}
\end{figure} 

An imbalance of flux between positive and negative polarities is observed during the decay phase of several active regions, for example, 11184 and 11512 in Figure 2. We explored whether this imbalance was due to the follower polarity decaying more rapidly with the result of dispersed flux becoming weaker than our threshold of 575 Mx cm$^{-2}$. Another possibility may be that the follower polarity flows out of the field of view.  At a given time, the flux imbalance of 11184 (11512) was 35\%~(14\%), respectively. By lowering the threshold to 220 Mx cm$^{-2}$, the flux imbalance was reduced to 18\%~(9\%). However, lowering the threshold further to 100 Mx cm$^{-2}$ only reduced the imbalance to 15\%~(7.5\%) with values not tending towards zero for zero threshold. It appears that, in cases where an imbalance of polarity exists, somewhat over half of the imbalance is due to the threshold not including the dispersed, decaying elements, while the rest is due to one polarity leaving the bounding box of the region. We note that the topic of flux imbalance in active regions has been researched by \cite{choudhary:2002}, \cite{chen:2012} and others who find that between one-third to one-half of active regions in their sample show a flux imbalance between 10-40\% of which some, but not all, can be attributed to instrumental effects. 


\begin{figure}[t!]
\centering
\includegraphics[scale=0.85]{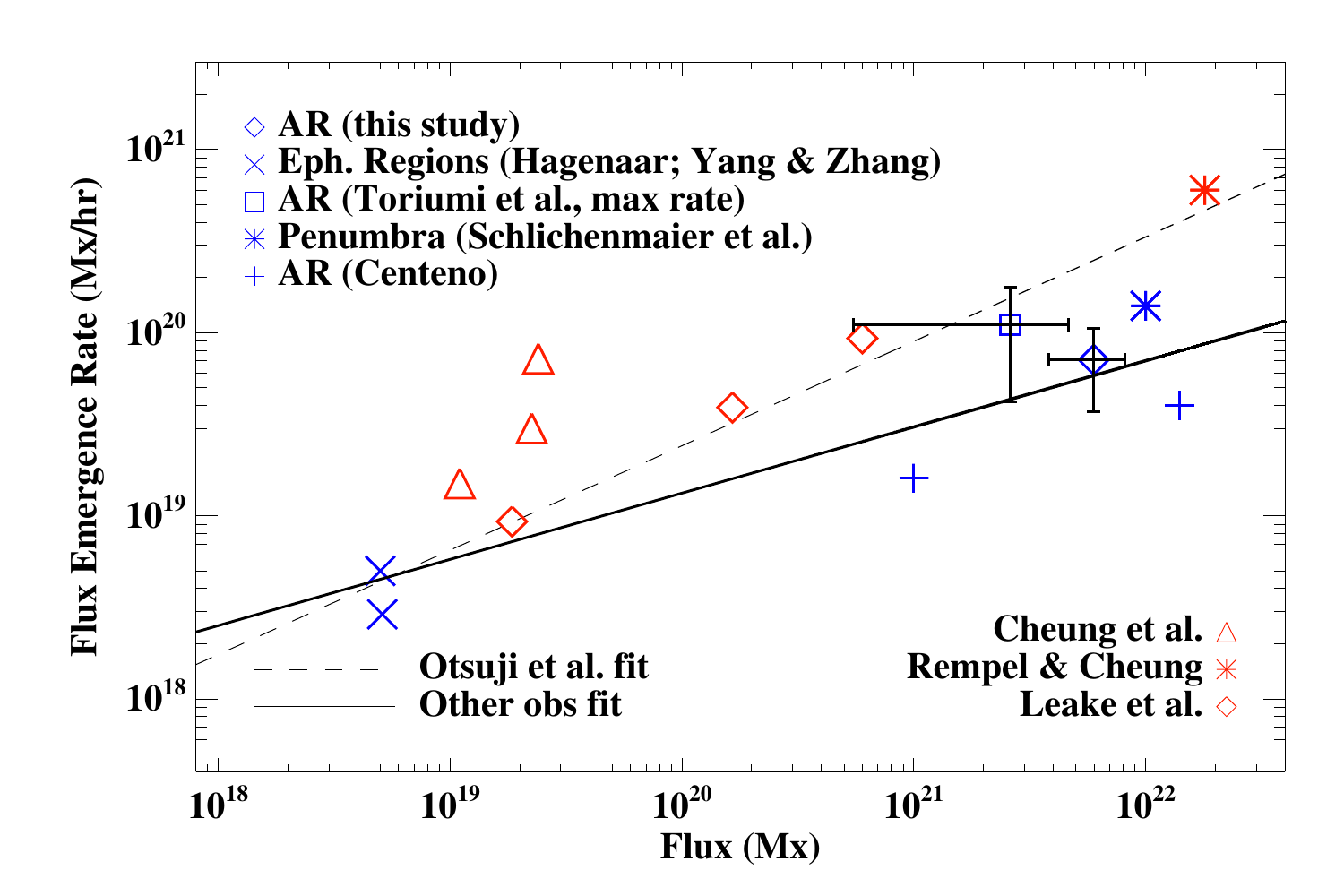}
\caption{Signed flux emergence rates (Mx hr$^{-1}$) as a function of peak magnetic flux (Mx) at the photosphere. The average flux emergence rate found in this study of ten active regions (blue $\diamond$, the average of p and f values shown in Table 3) is compared with modeling results (red $\bigtriangleup, \large *, \diamond$) and observational studies of ephemeral region (blue $\times$), active region (blue $+, \square$) and penumbra ($\large *$) emergence rates reported in the literature. The precise flux and emergence rate values plotted here as well as the reference papers in which they were published are summarized in Table 4 as $\phi$ and $\tau_e$. We report that the emergence rate scales as a power law dependent on peak flux values with an exponent of 0.36 (i.e., $\propto\phi_{max}^{0.36}$) (solid line). \cite{otsuji:2011} report a flux emergence rate with a power law dependent on peak flux whose exponent is 0.57 (i.e., $\propto\phi_{max}^{0.57}$) (dashed line).  }
\label{figure-area-emergence-log}
\end{figure} 

We now explore the hypothesis that flux emergence and decay rates depend on the peak flux in the active region. Figure \ref{figure-area-decayplots} shows emergence (left, $d\phi$/$dt$) and decay rates (right, $-d\phi$/$dt$) versus the peak flux of that polarity. Within our selection of sunspots, with a quite limited range of flux values, it is not convincing that emergence rate depends on peak flux. We tested a linear and power law relationship for both emergence and decay. The linear Pearson correlation coefficients for emergence rates were 0.09 (0.28) for p (f) spots and an exponent of zero best fit the log-log emergence data for all spots, meaning no correlation was present.

Decay rates were better correlated with a linear Pearson correlation of $0.78$ ($0.63$) for p (f) spots. Decay rates, as fit for both the p and f spot data, scaled with peak flux as a power law with exponent of 0.57 so that $- d{\phi}/dt \propto {\phi}_{max}^{0.57}$, see Figure 4, right panel. Active region decay rates have been observed to be proportional to the square root of the sunspot area \citep{petrovay:1997} and explained as an erosion of the perimeter of the sunspot by flows and turbulence, i.e., $- d{\phi}/dt \propto -dA/dt \propto A^{0.5}$ where A is the sunspot area. Assuming $\phi = \vec B \cdot A $ and a constant field strength, one would expect that $- d{\phi}/dt \propto \phi^{0.5}$. However, a larger sample with a wider range of flux values would be necessary to confirm this observed power law relationship.

We further investigate the dependence of emergence rates on peak flux, see Figure 5, when we plot the mean values determined within this study in the context of the greater literature.  With a wider range of flux values from ephemeral regions to large active regions, $10^{19}-10^{23}$ Mx, observed by Hagenaar (2001), Centeno (2014), Yang and Zhang (2014), Schlichenmaier et al. (2010) as well as the more comprehensive studies by Otsuji et al (2011) and Toriumi (2014), the emergence rate shows a more convincing picture of a possible dependence on peak flux.  The precise flux and emergence rate values plotted in Figure 5, as well as the reference papers in which they were published, are summarized in Table 4 as $\phi$ and $\tau_e$.

The blue (red) symbols in Figure 5 represent observed (modeled) values.  Ephemeral region flux values are generally considered to be within the range of $3 - 100 \times 10^{18}$ Mx, so the six points on the left side of Figure 5 would be considered ephemeral regions.  

In order to fit the observations, including average values from this paper as well as those values reported by others, we consider all the blue points in Figure 5 except the Toriumi et al. (2014) measure. Toriumi et al. (2014) was excluded because they reported the maximum rate of flux emergence instead of an average rate. We assume the flux emergence rate depends on peak flux as, $d{\phi}/dt = A \tilde{\phi}_{max}^{c}$.   The quantities $d{\phi}/dt$ and $A$ are in units of Mx hr$^{-1}$. The quantity $\tilde{\phi}_{max}$ is the maximum, or peak, flux, $\phi_{max}$, normalized by a unit Mx flux in order to have a unitless quantity. The dependence was determined using the method described by \citet{fisher:1998} in which they explore the luminosity dependence on flux.  This was done by computing the Spearman correlation coefficient $r_{s}$ between $\tilde{\phi}_{max}$ and $A= d{\phi}/dt / \tilde{\phi}_{max}^c$ for various powers $c$.  The power $c$ is varied until the correlation vanishes $(r_{s} = 0)$. We then have two variables, $A$ and $c$. As a result, the dependence of emergence rate can be determined. The final fit was found to be $d{\phi}/dt = 7.8 \times 10^{11} \tilde{\phi}_{max}^{0.36}$.

In comparison, \cite{otsuji:2011} report a dependence as, $d{\phi}/dt = 9.6 \times 10^{7} \phi_{max}^{0.57}$ where the quantities $d{\phi}/dt$, $\phi_{max}$ and $9.6\times10^{7}$ are in units of Mx hr$^{-1}$, Mx, and unspecified, respectively.  It is unclear how the constant, $9.6 \times 10^{7}$ was determined, but they derived an analytical relationship between the maximum flux and flux growth (or emergence) rate to be $\propto\phi_{max}^{0.50}$ assuming that the emergence occurs in discrete $\Omega$ loop sizes (4 Mm) whose rise velocity and aspect ratio is constant just beneath the photosphere. 

Also overplotted in Figure 5 are several rates determined from simulations by Cheung et al. (2007, 2008), Rempel and Cheung (2014) and Leake et al. (2017, in preparation).  A relationship is clear in the simulations, showing that higher flux concentrations emerge faster.  

The two red triangles representing fluxes of $2.25 \times 10^{19}$ and $1.1\times 10^{19}$ Mx, with associated emergence rates of $3 \times 10^{19}$  and $1.5 \times 10^{19}$ are results from Cheung et al. (2008) and show that given similar conditions, a flux rope with a larger twist will emerge flux faster as the twist parameter is the only difference between these two data points.  Simulations by Nozawa (2005) and Murray et al. (2006), and others, also show that sheared or twisted flux tubes emerge faster than flux tubes without shear or twist.  

In the study of Leake et al. (2017), whose results are depicted as red diamonds in Figure 5, the free parameters of the initial idealized convection zone magnetic flux are varied, namely the magnetic field strength, radius, initial depth, and ratio of axial to azimuthal flux (or twist). These four parameters determine the emergence rate in a complex way. Leake et al. (2016) varies the axial magnetic flux (which corresponds closely to the unsigned flux upon emergence) by changing the field strength, radius and initial depth such that the initial rise speed and ratio of radius to depth are kept constant for the three different simulations shown in Figure 5. The twist of the flux rope, which positively affects the emergence rate, is assigned as the smallest value that results in a minimum of 50\% of the initial flux emerging above the model surface. This can be seen in the fourth and fifth column of Table 4 where the Leake et al. entries show half of the axial magnetic flux assigned at the base of the simulation, $\phi_b$, emerges as flux in the photosphere, $\phi$. 

The results from the Leake et al. (2017) simulations correlate surprisingly well with the Otsuji et al. (2011) fit, shown in Figure 5.  This agreement was not in any way superficially forced. The maximum rate of flux emergence observed by Toriumi et al. (2014) also lies perfectly along the Otsuji et al. (2011) fit. It may be that the Centeno (2011) and the Otsuji et al. (2011) results represent a lower and upper bound, respectively, for observed active region flux emergence rates and that any rates in between are within solar range.  The two outlier points from Rempel \& Cheung (2014) and Cheung et al. (2008) may be higher than the observed solar range because the convective flows in the models are too high. 

It is unclear how the assigned flux geometry (i.e., torus versus slab) at the bottom boundary of the simulations affects the emergence rate. \cite{hood:2009} conducted flux emergence simulations using toroidal loops and found that the plasma drainage along the tube is effective and that for large values of field strength, the axial field of the toroidal loop fully emerges into the atmosphere.

\begin{table}[t!]
\footnotesize
\centering
\begin{tabular}{llccccc}
\hline\hline
Author(s) & Year & Type & $\phi_b$ (Mx) & $\phi$ (Mx) &  $\tau_e$(Mx hr$^{-1}$)& Ref\\ 
\hline
Cheung et al. & 2008 & Sim & 1.55 $\times~10^{20}$&1.1 $\times~10^{19}$&  1.5 $\times~10^{19}$ & ApJ, 687, 1373\\ 
Leake et. al& In prep &Sim  &  3.7 $\times~10^{19}$&1.85 $\times~10^{19}$&  9.3 $\times~10^{18}$ & In preparation\\ 
Cheung et al. & 2008 &Sim  &  1.55 $\times~10^{20}$&2.3 $\times~10^{19}$&  3 $\times~10^{19}$ & ApJ, 687, 1373\\ 
Cheung et al. & 2007 &Sim  &  8.5 $\times~10^{18}$ &2.4 $\times~10^{19}$&  7.2 $\times~10^{19}$ & A\&A, 467, 703\\ 
Leake et. al& In prep &Sim  &  3.3 $\times~10^{20}$&1.65 $\times~10^{20}$&  3.9 $\times~10^{19}$ & In preparation\\ 
Leake et. al& In prep &Sim  &  1.2 $\times~10^{21}$&6.0 $\times~10^{20}$&  9.3 $\times~10^{19}$ & In preparation\\ 
Rempel \& Cheung & 2014 &Sim  & 3.6 $\times~10^{22}$&1.8 $\times~10^{22}$ &  6 $\times~10^{20}$ & ApJ, 785, 1\\ 
\hline
Yang \& Zhang & 2014 & Obs &-& 5.0 $\times~10^{18}$ (ephemeral) & 5.0 $\times~10^{18}$& ApJ, 781, 7\\
Hagenaar & 2001 &Obs &- &  5.1 $\times~10^{18}$ (ephemeral)&  2.9 $\times~10^{18}$ & ApJ, 555, 448\\ 
Otsuji et al. & 2011 &Obs  &-& 4.1 $\times~10^{20} $(mean)&  3.0 $\times~10^{19}$ &PASJ, 63, 1047\\ 
Centeno & 2012 &Obs &-& 1.0 $\times~10^{21}$ (AR11211)&  1.6$\times~10^{19}$ & ApJ, 759, 72\\ 
Toriumi et al. & 2014 &Obs&- & 2.6 $\times~10^{21} $(mean) & 1.1 $\times~10^{20}$ &ApJ, 794, 19\\
Norton et al. & 2016 &Obs &- & 5.9 $\times~10^{21} $(mean)&  7.1 $\times~10^{19}$ & This paper\\ 
Schlichenmaier et al. & 2010 & Obs&- & 1.0 $\times~10^{22}$ (penumbra)& 1.4 $\times~10^{20}$ & A\&A, 512, L1\\
Centeno & 2012 &Obs&-& 1.4 $\times~10^{22}$ (AR11105)&  4 $\times~10^{19}$ & ApJ, 759, 72\\ 
\hline
\end{tabular}
\caption{\small A summary of flux emergence rates found in the literature from both simulations and observations.  The columns correspond to primary author(s), year of publication, whether it is a simulation or observation, longitudinal flux assigned at the base of the simulations ($\phi_b$) in the case of simulation entries, peak signed flux at the photosphere, signed flux emergence rate and reference. Simulations comprise the top rows while observations are shown below. The photospheric fluxes, $\phi$, are listed from smallest to largest from top to bottom within the simulation and observation sections. The Cheung et al. (2007) entry is the mean of eight simulation runs.}
\label{table-sim-obs-compare}
\end{table}

In Figure 6, we show two examples of observed active region emergence (AR 11428 and 11460) and decay compared to one example from numerical simulations \citep{rempel:2014}. The simulated flux emergence rate is much faster.  In terms of absolute time of AR emergence, the sunspot groups finished emerging within 25$-$165 hours of pore production, with a mean time at the emergence ending at 78 hours. This agrees well with \cite{harvey-angle:1993} who examined 978 active regions and concluded that active regions emerged within five days or less and that the emergence time was short compared to the lifetime of the active region.  It is informative to simply examine the shape of the flux emergence and decay curves and compare these to simulations, as shown in Figure 6.  The simulations show a lag time between the flux emergence (blue curve, lower left panel) and the umbral formation (red curve, lower left panel). The observations do not show this. However, our observations shown in Figure 6 do not begin at a time where there is zero flux. Instead, we define the start of the flux emergence event as when a pore is visible, meaning some flux is already contained within the field of view and indeed within a dark intensity contour at the beginning of the observations. As such, our observations probably begin around 30 hours into the simulation of Rempel \& Cheung. It is worth noting that the simulations without convection do not produce simulated intensity maps, and so intensity cannot be used to distinguish between umbra, penumbra, and quiet-Sun, as we have done here for the observations and for the convection-based emergence simulations. 


%
\begin{figure}[t!]
\begin{minipage}[b]{0.48\textwidth}
\includegraphics[width=\textwidth]{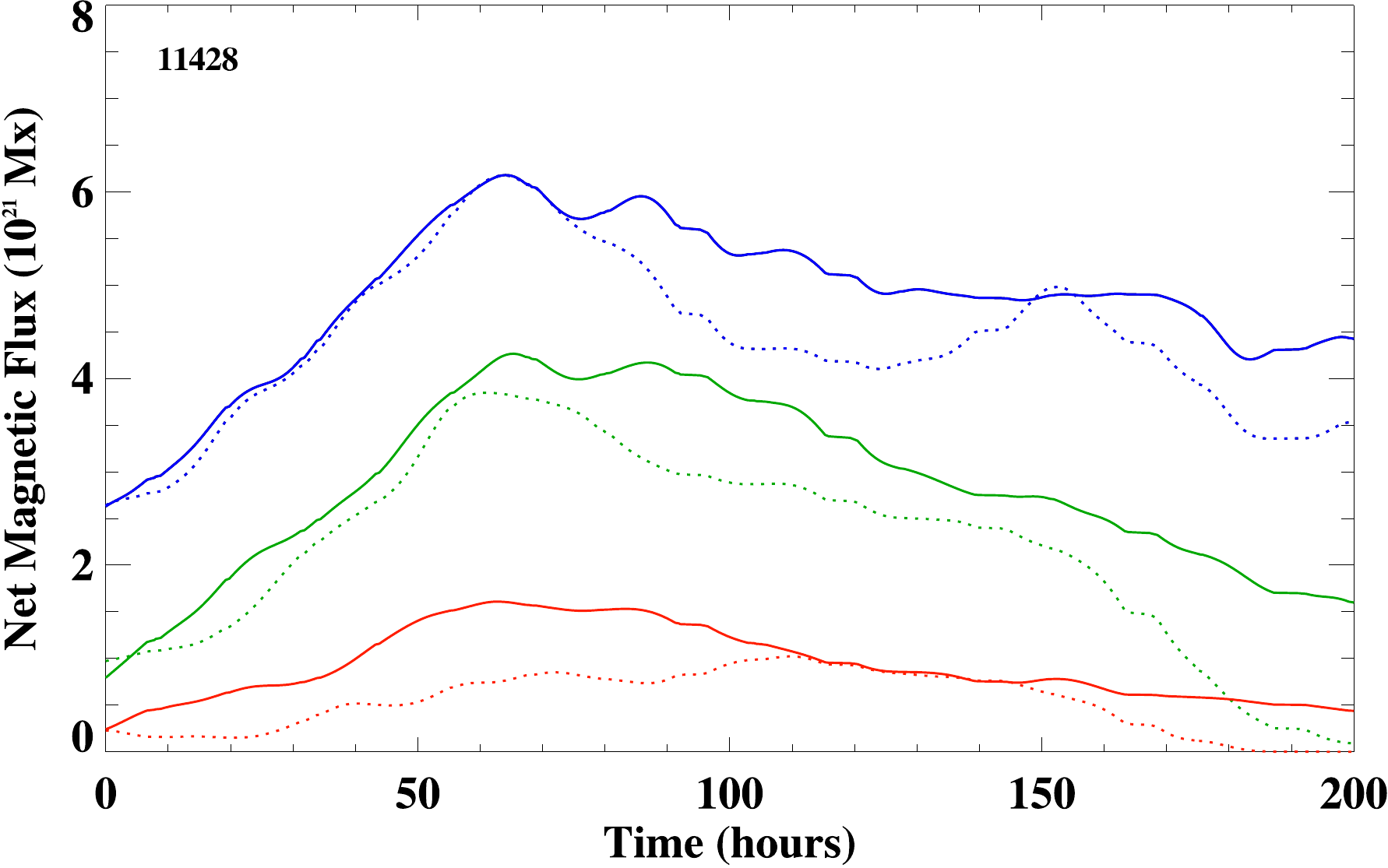}
\end{minipage}
\begin{minipage}[b]{0.48\textwidth}
\includegraphics[width=\textwidth]{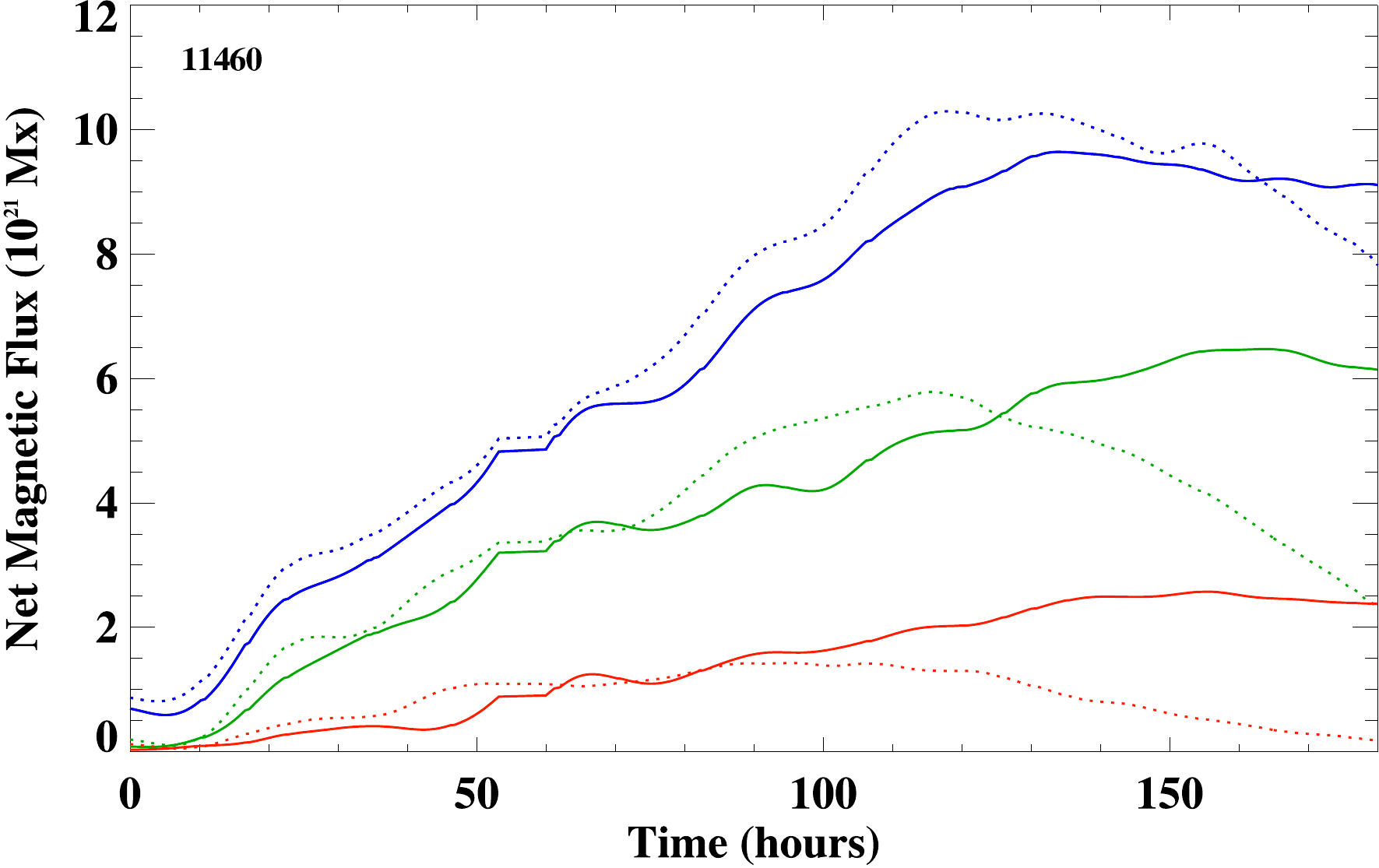}
\end{minipage}
\begin{minipage}[b]{0.48\textwidth}
\includegraphics[width=\textwidth]{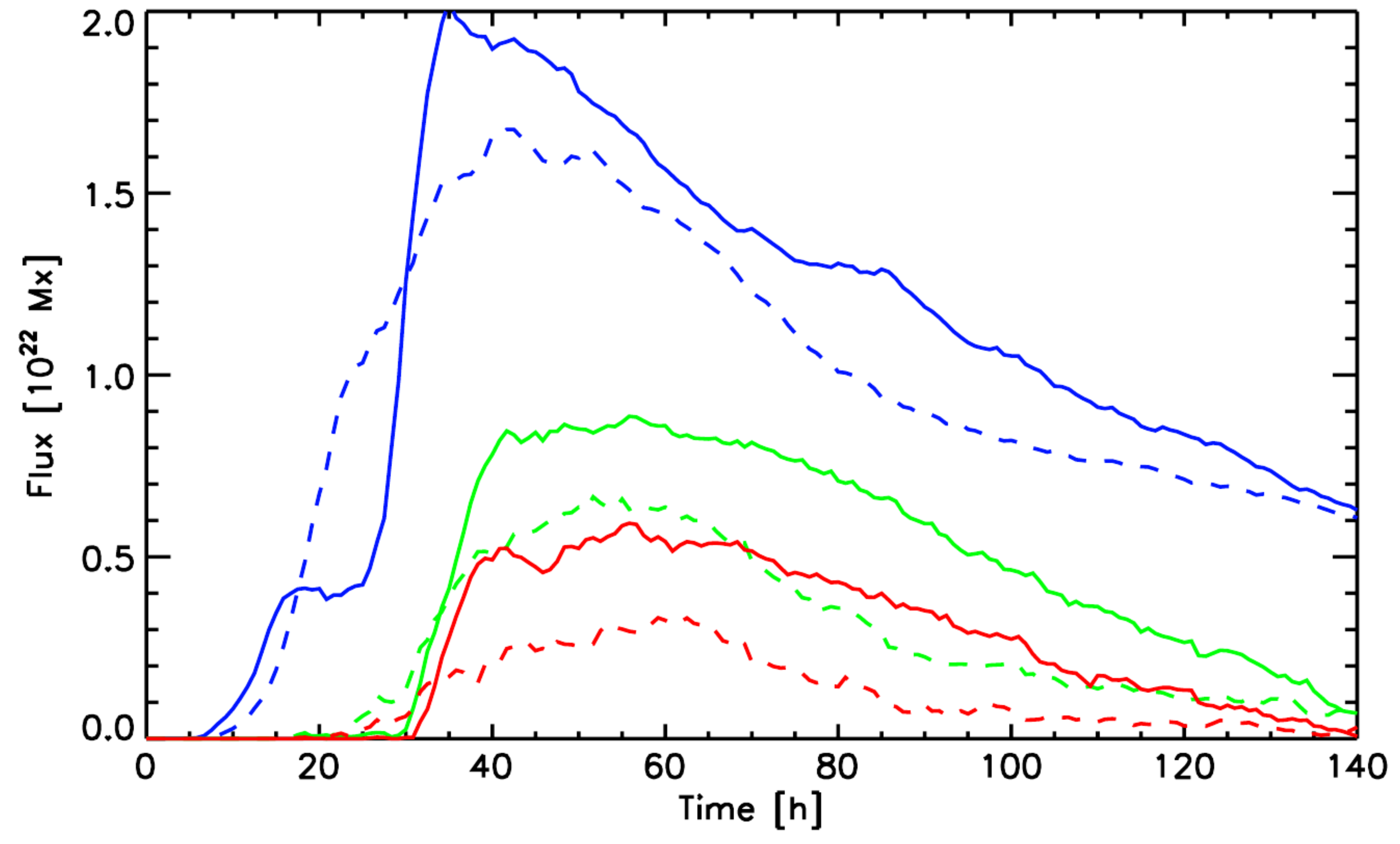}
\end{minipage}

\caption{Top: NOAA 11428 (left) and 11460 (right) with preceder (solid line) and follower (dashed line) unsigned patch flux represented by blue lines, unsigned sunspot flux represented by green lines, and unsigned umbra flux represented by red lines.  The top two panels are also seen in Fig 1 and 2. Bottom: A figure reproduced with permission from \cite{rempel:2014}. Here, blue lines represent the unsigned flux separated by polarity (but with no intensity threshold), green lines represent flux within intensity contours defined by $\bar I$ $<$ 0.8 $I_{\odot}$, and red lines represent flux within intensity contours defined by $\bar I$ $<$ 0.5 $I_{\odot}$. The flux emergence and decay curves, including the overall shape and behavior, can easily be compared between simulations and observations.  }
\label{figure-rnc}
\end{figure}

Comparing the observed flux emergence behavior with that of simulations in Figure 6, we observe emergence to occur for the NOAA 11428 and 11460 active regions over the course of 3-5 days whereas the simulation shows the flux emergence occurs in less than 2 days. The \cite{rempel:2014} simulation shown in Figure 5 contained flux higher than any in our observations, so perhaps some of the difference in observations and simulations can be accounted for by the difference in flux. We hypothesize that there are several reasons why the flux emergence rates produced in simulations are faster than observed.  The flux emergence rates are too fast due to the initial rise speed being too large. The rise speed is set in two ways, either by too great a buoyancy or too great an advection speed across the bottom boundary. For example, a value of 500 m s$^{-1}$ is used by \cite{rempel:2014} to advect flux across the bottom boundary located at 15.5 Mm and this contributes to the fast emergence in the simulations. A recent paper by \cite{birch:2016} reports an upper limit on the rise speed of 150 m s$^{-1}$, based on helioseismology measurements, for active region magnetic flux concentrations at a depth of 20 Mm below the photosphere. 
 
An interesting feature absent in the modeling results is the plateau of total flux at the end of emergence, see , i.e. AR 11141, 11184 11682 in Figure 2. The stability can last a day or two. It may be absent in the modeling results shown in Fig 6c since Rempel \& Cheung (2014) opened up the the lower boundary roughly twenty hours into the simulation. The plateau in the observations could indicate that at whatever depth the sunspot magnetic fields are rooted, they do not allow convective intrusions and can maintain their stability for a few days.

\section{Conclusions}

We report on a single, easily measured quantity --  the magnetic flux emergence rate of active regions in the mid to lower photosphere -- hoping that the value and its dependence on peak flux can better inform us of the processes by which flux emerges from the near surface layers into the solar atmosphere. The rates and flux values are determined from HMI continuum intensity and vector magnetogram data. 

 The ten bipolar regions examined in this paper contain an average signed flux of $5.9\times$10$^{21}$ Mx emerging at 7.1$\times$10$^{19}$ Mx h$^{-1}$ while decaying at half of that rate, see Table 3. These values are determined without using intensity thresholds, i.e. they contain plage and sunspot flux, but after applying a lower threshold of 575 Mx cm$^{-2}$ to minimize the contributions of noisy transverse field measurements.  As such, we interpret the rates and fluxes as lower limits.  The ten regions show considerable scatter in flux emergence rates with a standard deviation of 3.5 $\times$ 10$^{19}$ Mx h$^{-1}$, half of the average rate. 

The average flux decay rate of 3.2 $\times$ 10$^{19}$ Mx h$^{-1}$ is half that of the emergence rate, see Table 3. This decay rate is in reasonable agreement with the rate at which moving magnetic features are expected to carry away magnetic flux.  See, for example, the work done by \cite{hagenaar:2005}, who found moving magnetic feature (MMF) transport rates of $(0.4-6.2) \times $10$^{19}$ Mx h$^{-1}$. 

The small range of flux values sampled with HMI in this paper do not result in any correlation between emergence rates and peak flux, see Figure 4 left panel. However, when examined in context with other reports of flux emergence rates, see Table 4, we see a clear trend that higher flux regions emerge faster and the rate is dependent on the total flux of that region.  Using a synthesis of values found in the literature, we find flux emergence rates to scale as a power law dependent on peak flux values such that
$d{\phi}/dt = 7.8 \times 10^{11} \tilde{\phi}_{max}^{0.36}$ where the quantities $d{\phi}/dt$,  and $7.8\times10^{11}$ are in units of Mx hr$^{-1}$ and $\tilde{\phi}_{max}$ is normalized to be unitless.  In contrast, Otsuji et al. (2011) reports the emergence rates scale as a power law with an exponent value of 0.57 as observed with Hinode SOT.  The difference in power law exponents reported herein and by Otsuji et al. (2011) is not due to the thresholding of HMI data, and likely due to the fitting of a composite set of values taken by multiple instruments over longer durations (days) compared to a single instrument observing regions for a shorter duration (hours).

Numerical modelling results also clearly show that higher flux regions emerge faster. However, the observed values of $d{\phi}/dt$ reported herein are smaller than those reported in simulation results by Rempel \& Cheung (2014) and others. This may indicate a slower rise of the magnetic flux in the solar interior than what is captured or assigned in the simulations. 

\section*{Acknowledgement}
This work was supported by LWS under ROSES NNH13ZDA001N to the U.S. Naval Research Laboratory and by NASA Contract NAS5-02139 (HMI) to Stanford University. Computational resources were provided for J. E. Leake and M. G. Linton by the Department of Defense High Performance Computing Program.

\bibliographystyle{apj}
\bibliography{paper}
\end{document}